\documentclass[12pt,aps,prb,preprint]{revtex4-1}   

\usepackage{amsmath}    
\usepackage{graphicx}   

\begin{document}

\title{Enhancing the understanding of entropy through computation}
\author{Trisha Salagaram}
 \affiliation{University of Pretoria,Physics Department, Pretoria, 0001, South Africa}
\author{Nithaya Chetty}
 \altaffiliation{National Institute for Theoretical Physics, Johannesburg, 2000, South Africa}
 \affiliation{University of Pretoria,Physics Department, Pretoria, 0001, South Africa}
 \email{trisha.salagaram@up.ac.za}   
\date{\today}

\begin{abstract}
We devise a hierarchy of computational algorithms to enumerate the microstates of a system comprising $N$ independent, distinguishable particles. An important challenge is to cope with integers that increase exponentially with system size, and which very quickly become too large to be addressed by the computer. A related problem is that the computational time for the most obvious brute-force method scales exponentially with the system size which makes it difficult to study the system in the large $N$ limit.  Our methods address these issues in a systematic and hierarchical manner. Our methods are very general and applicable to a wide class of problems such as harmonic oscillators, free particles, spin $J$ particles, etc. and a range of other models for which there are no analytical solutions, for example, a system with single particle energy spectrum given by $\varepsilon(p,q)\,=\,\varepsilon_0(p^2+q^4),$ where $p$ and $q$ are non-negative integers and so on. Working within the microcanonical ensemble, our methods enable one to directly monitor the approach to the thermodynamic limit ($N\rightarrow \infty)$, and in so doing, the equivalence with the canonical ensemble is made more manifest. Various thermodynamic quantities as a function of $N$ may be computed using our methods; in this paper, we focus on the entropy, the chemical potential and the temperature.
\end{abstract}

\maketitle

\section{Introduction}
Entropy is critical to the understanding of statistical physics. However, it is our experience that students have difficulty in conceptualizing entropy. For example, the following simple problem has been presented to introductory statistical physics students at our institutions\cite{Institution} on numerous occasions over more than a ten year span, and the question has usually been bewildering to many of them:

\vspace{0.5cm}
\noindent {\em Imagine a container with a large number of red marbles and another container with a similar number of green marbles. Mix the marbles together, and shake the container really well. After this, a handful will draw, on average, an equal number of red and green marbles. Question: How does this come about - do the red marbles 'know' of the green marbles, and vice versa?!}
\vspace{0.5cm}

Of course the red marbles do not communicate with the green marbles! The answer has to do with the fantastically large number of ways in which the system can configure itself for which the final state may be regarded as a mixed state compared to a separated or even partially separated state.

Computation has a significant role to play in enhancing the understanding of entropy. Entropy, as in the above example, is often presented in the context of the second law of thermodynamics as a measure of the state of randomness of a system. The Monte Carlo method, therefore, is a very obvious computational method in the study of entropy, and many pedagogical texts focus on the use of the random variable to simulate various thermodynamic processes.\cite{mc1,mc2,tobochnik1,tobochnik2}

In this paper, however, we approach entropy by computing the relevant statistical quantities more directly. Our starting point is the enumeration of system microstates. This very quickly becomes a computational challenge because we have to cope with integers that increase exponentially with the system size, and therefore become larger than the largest integer that can be addressed by a computer. A first step in our hierarchy of computational methods to address this problem is to work with real numbers using real arithmetic, but this improvement is minuscule compared with the scale of the problem. A related problem is that the computational time for the most obvious brute-force method scales exponentially with the system size which makes it difficult to study the system in the large $N$ limit.

We devise a hierarchy of computational algorithms to enumerate the microstates of a system comprising $N$ independent, \underline{distinguishable} particles. Our methods are very general and applicable to a wide class of problems such as $\varepsilon(p)=\varepsilon_0 p$ (quantum harmonic oscillators), $\varepsilon(p)\,=\,\varepsilon_0 p^2$ (1D free particles),  $\varepsilon(p,q,r)\,=\,\varepsilon_0(p^2+q^2+r^2)$ (3D free particles), spin $J$ particles with $\varepsilon(m)=\varepsilon_0 m$ where $m=-J, -J+1,\cdots,J$, etc. and also a range of other models for which there are no analytical solution, for example, $\varepsilon(p)\,=\,\varepsilon_0 p^3$, $\varepsilon(p,q)\,=\,\varepsilon_0(p^2+q^4)$, etc. Here $p$, $q$ and $r$ are integers. $J$ is integral or half odd integral for real spin systems. The systems may be physically realizable or they may be idealized. They may or may not be solvable analytically. These models may have a finite or an infinite single particle energy spectrum. Our numerical method is applicable to all of these systems and has been tested against various analytically solvable models such as a system of harmonic oscillators, a system of free particles, a system of spin $1$ particles, etc. This has enabled us to check the accuracy of our numerical methods.
Working within the microcanonical ensemble, our methods enable one to directly monitor the approach to the thermodynamic limit ($N\rightarrow \infty)$, and in so doing, the equivalence with the canonical ensemble is elucidated. Various thermodynamic quantities as a function of $N$ may be computed using our methods; in this paper, we focus on the entropy, the chemical potential and the temperature.

Furthermore, our work elucidates different computational algorithms that give the student an experience of different ways of solving the same problem, with some methods being exponentially more efficient than others. This point is worth pondering about: We do not seem to be spending sufficient time focusing students' minds on the important but distinct aspect related to algorithm development in the teaching of computational physics. Very often, we go from discussions of a physics problem directly to code development, often skimping on the details of algorithm development. Our presentation in this paper encourages the student to think a little more carefully about the manner in which the problem should be solved, and gets the student to appreciate how a poorly thought algorithm, even though mathematically correct, can lead to catastrophe.

In our work, temperature as a statistical quantity emerges very naturally. For small system sizes, the curve of entropy versus energy is jagged and thus not differentiable, which makes the temperature of the system ill-defined since
\begin{equation}
\label{Temperature}
\frac{1}{T}=\left(\frac{\partial S}{\partial E}\right)_{NV}.
\end{equation}

As the large system size limit is approached, we show that the curve of entropy versus energy converges to a smooth, differentiable form for which temperature is numerically better behaved; this is consistent with the fact that temperature for an isolated system only really makes sense in the thermodynamic limit. Our approach gives a clear indication of the rate at which the thermodynamic limit is reached, and what 'large' actually means from a computational point of view.

The subject of the enumeration of states is, of course, an established one\cite{Pathria} that has been developed since the time of Boltzmann, more than a hundred years ago. However, publications over the course of the past ten years\cite{Moore,Prentis,Schoepf,Pena} show that the pedagogical understanding of the subject is still in need of attention, and whilst the theoretical concepts are well understood, the numerical evaluation of results is still lagging behind arguably for the reasons given above, namely, the numbers increase too rapidly which make them difficult to handle computationally.

Analytical expressions for the number of system microstates, namely, $\Omega(N,E)$ for $N$ particles with total energy $E$,  only exist for a small number of problems such as a  system of $N$ independent quantum harmonic oscillators or a system of $N$ free particles. Here the Stirling approximation\cite{Stirling} is applicable and enables the derivation of closed form analytical results. For a wide range of other problems, such analytical solutions do not exist, and accurate numerical methods are needed.

Moore and Schroeder\cite{Moore} considered a system of quantum harmonic oscillators and noted that the results become 'rather cumbersome with much more than  a few hundred energy units, while [computer] overflow errors can occur if there are more than a few thousand oscillators ...'. We present in this paper a robust means of avoiding this type of overflow error. Prentis and Zainiev\cite{Prentis} considered systems in the large $N$ limit. In our work, we compute $\Omega(N,E)$ recursively for $N=1,2,3 \cdots N_{max}$ where $N_{max}$ can be made sufficiently large to test various results in the large $N$ limit, i.e using our method, we are able to approach the thermodynamic limit systematically and with no approximations. Our algorithm is numerically stable, and the results are achievable in real-time on a moderate modern desk-top computer.\cite{computer} Schoepf\cite{Schoepf} considered equally spaced energy levels in studying entropy and the approach to thermodynamic equilibrium, and remarked that 'students have a better intuitive feel for the concept of energy, an intuition that is lacking for entropy', a point that we strongly agree with.

In Section II, we present a non-trivial model for which no analytical results exist. We discuss a hierarchy of methods that enable a complete and numerically exact resolution of the problem. Working within the microcanonical ensemble, we give results for the entropy and we compare our results as a function of $N$ with the equivalent results in the canonical ensemble which are, by definition, in the thermodynamic limit. This enables us to directly monitor the systematic approach to the thermodynamic limit and to test the equivalence of the two ensembles. This has enormous pedagogical value. We consider the temperature and the chemical potential, and we make similar comparisons with equivalent results in the infinite size limit. In Section III, we draw our conclusions and give suggestions for further work.

\section{The counting of microstates}

\subsection{The model}
As indicated in the introduction, our methodology is very general and applicable to a very wide range of different problems. Here we have chosen to focus on the following model because it is not analytically solvable and because it has non-trivial structure. This model is more complex than the analytically solvable models. We consider an isolated system consisting of $N$ identical, non-interacting quantum particles with single-particle energy given by
\begin{equation}
\label{energy}
\varepsilon(p,q)=\varepsilon_0\left(p^2 + q^4\right),
\end{equation}
where $p$ and $q$ are non-negative integers. We wish to determine the total number of system microstates $\Omega(N,E)$ accessible to the system with energy $E$ and hence the entropy given by \begin{equation}\label{defn_entropy}S(N,E)=k_B\ln\Omega(N,E),\end{equation} where $k_B$ is the Boltzmann constant. $\Omega(N,E)$ depends on whether the particles are distinguishable or not, and whether they are bosons or fermions.
\begin{figure}[t]
\begin{center}
\scalebox{0.5}{\includegraphics{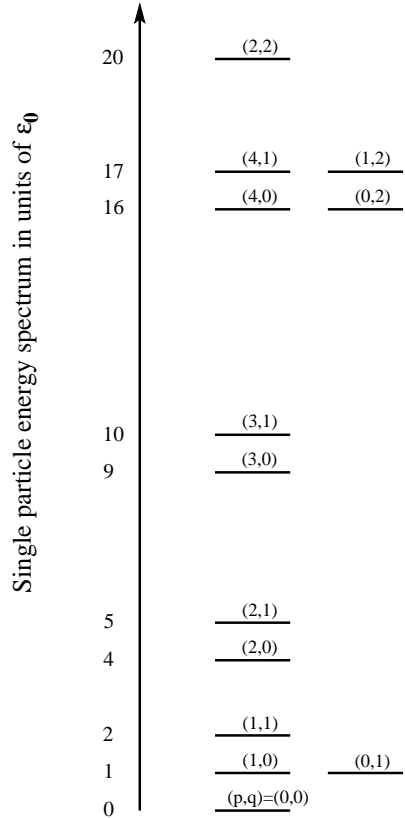}}
\caption{\label{SingleParticle} The single particle energy spectrum $\varepsilon(p,q)=\varepsilon_0(p^2+q^4)$ where $p$ and $q$ are non-negative integers.}
\end{center}
\end{figure}
Figure \ref{SingleParticle} shows the single particle spectrum.  This model is non-trivial because of the presence of degeneracies, for example, $\varepsilon(1,0)=\varepsilon(0,1)=\varepsilon_0$, and because the energy levels are not simply spaced. This makes computing $\Omega(N,E)$ a non trivial task. Later, to show the versatility of our method, we also present results for spin $1$ particles and for a system of quantum harmonic oscillators.

\subsection{'Pen-on-paper' solution}
To start with, it is useful to list the system microstates for a small number of particles. For example, for $N=3$, and $E=20\varepsilon_0$, the microstates can easily be listed. Let $(p_i,q_i)$ be the quantum labels for the $i^{th}$ particle, then the system microstates $(p_1,q_1;p_2,q_2;p_3,q_3)$ for indistinguishable spinless particles (bosons) are (2,2;0,0;0,0), (1,2;1,1;0,1), (1,2;1,1;1,0), (4,1;1,1;0,1), (4,1;1,1;1,0), (4,0;2,0;0,0), (4,0;1,1;1,1), (0,2;2,0;0,0), (0,2;1,1;1,1), (3,1;3,1;0,0), (3,1;3,0;0,1), (3,1;3,0;1,0), (3,1;2,1;2,1), (3,0;3,0;1,1), i.e. there are 14 microstates for indistinguishable particles. It can be easily inferred that there are 66 microstates for distinguishable particles, and 76 microstates for indistinguishable spin $\frac{1}{2}$ particles (fermions). Going through this exercise leaves one with the distinct impression that it is very cumbersome to try to solve this problem using 'pen-on-paper' for a large number of particles. An efficient computational algorithm is needed.

\subsection{A brute force computational method}
\begin{figure}[t]
\begin{center}
\scalebox{0.50}{\includegraphics{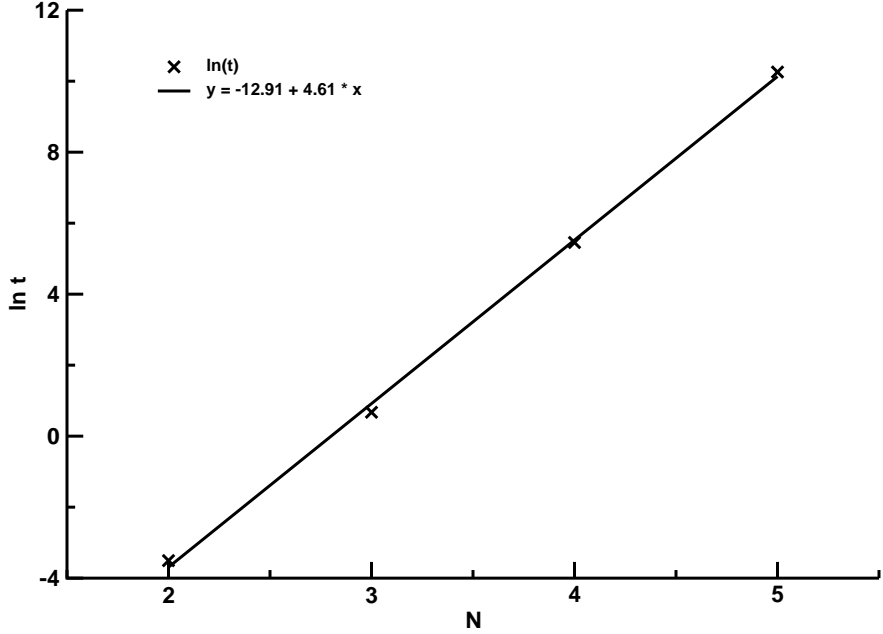}}
\caption{\label{lnTvslnN}Graph of $\ln t$ versus $N$, where $t$ is the real computational time in seconds and $N$ is the number of particles. In this graph the data points are plotted for $N = 2, 3, 4$ and $5$ particles.}
\end{center}
\end{figure}
For distinguishable particles, the simplest brute force method that can be devised involves $N$ nested {\em do}-loops, each over the list of single particle energy levels. This results in a computational scheme that scales exponentially with the system size. Still, this is an instructive method to apply because it gets the student to appreciate the rapid increase in $\Omega(N,E)$ for moderate $N$, and the corresponding exponential increase in computational time. Figure \ref{lnTvslnN} is a graph of $\ln t$ versus $N$, where $t$ is the real computational time\cite{computer} taken to compute $\Omega(N,E)$ for $N=2,3,4\, \mbox{and}\, 5$ and for $E=500\varepsilon_0$. The graph shows that $t \propto \exp(\alpha N)$, where $\alpha=4.61$ is the slope of the curve. We conclude that it will take several million years to consider only a moderate number of particles! A more practical algorithm is needed.

\subsection{A recursive solution for $\Omega(N,E)$}
\begin{figure}[t]
\begin{center}
\scalebox{0.50}{\includegraphics{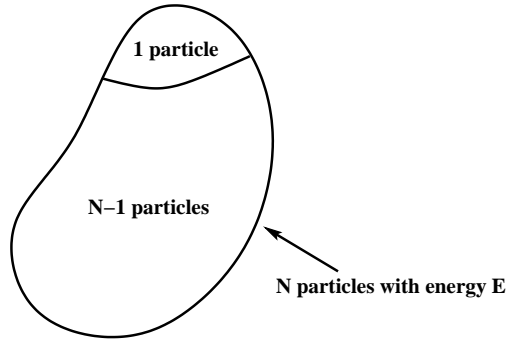}}
\caption{\label{System}The system comprising $N$ particles with energy $E$.}
\end{center}
\end{figure}

For fixed total energy $E$, we consider a system of $N$ distinguishable particles to be composed of a subsystem of $(N-1)$ particles with $\Omega(N-1,E-E^{'})$ microstates and a subsystem of $1$ particle with $\Omega(1,E^{'})$ microstates as depicted in Fig.~\ref{System}.
In this context, $\Omega(1,E^{'})$ are the degeneracies of the single particle levels with energy $E^{'}\leq E$. $\Omega(N,E)$ is then simply constructed by summing the number of ways in which the subsystem of $(N-1)$ particles can configure itself with energy $E-E^{'}$ weighted by the degeneracy of the single particle level with energy $E^{'}$, i.e.
\begin{equation}
\label{recursive}
\Omega(N,E)\,=\,\sum_{E^{'}= 0}^{E} \Omega(N-1, E-E^{'})\times\Omega(1,E^{'}).
\end{equation}
This is the basis of our recursive algorithm, which can be coded in a very straight forward manner. $\Omega(1,E^{'})$ can be constructed very simply and, from this, $\Omega(N,E)$ may be calculated recursively for $N=2,3,4, \cdots$.

\begin{figure}[t]
\begin{center}
\scalebox{0.50}{\includegraphics{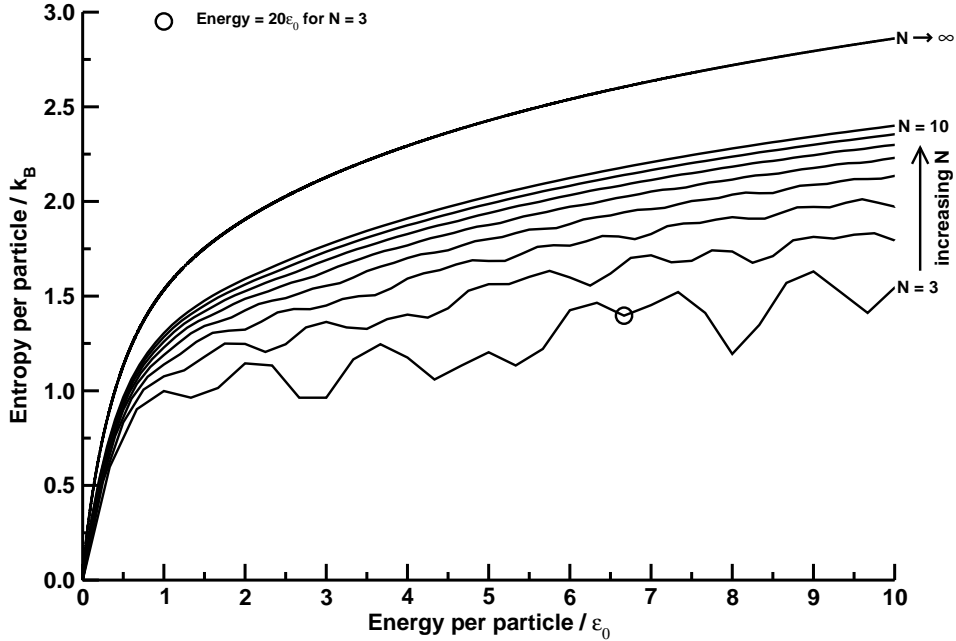}}
\caption{\label{SppvsEppN3to10}Entropy per particle versus energy per particle for $N=3 \-- 10$ distinguishable particles. The result in canonical ensemble ($N \rightarrow \infty$) is also included to illustrate where this system is relative to the thermodynamic limit ($N \rightarrow \infty$). The bulleted point depicts the calculated result of $\Omega(3,20\varepsilon_0)=66$ as discussed in subsection II B.}
\end{center}
\end{figure}
A plot of entropy per particle versus energy per particle is presented in Fig.~\ref{SppvsEppN3to10} for $N=3\--10$ particles, and compared with the equivalent result calculated within the canonical ensemble (see Appendix A) which is, by definition, in the thermodynamic limit. For these rather low values of $N$, the graphs are not smooth, which makes the temperature ill-defined. This is consistent with the fact that temperature for an isolated system only really makes sense in the infinite system size limit, a point already alluded to in the introduction in the context of Eq.~(\ref{Temperature}).

Using Eq.~(\ref{recursive}), we are able to compute $\Omega(N,E)$ for relatively large values of $N$. Using the computational resources at our disposal, the maximum number of particles that we were able to consider is $N\sim 200$. The computational time taken is of the order of minutes. For $N$ greater than this maximum number, the largest number addressable by the computer\cite{HugeTiny} is exceeded which results in overflow errors.

\begin{figure}[t]
\begin{center}
\scalebox{0.50}{\includegraphics{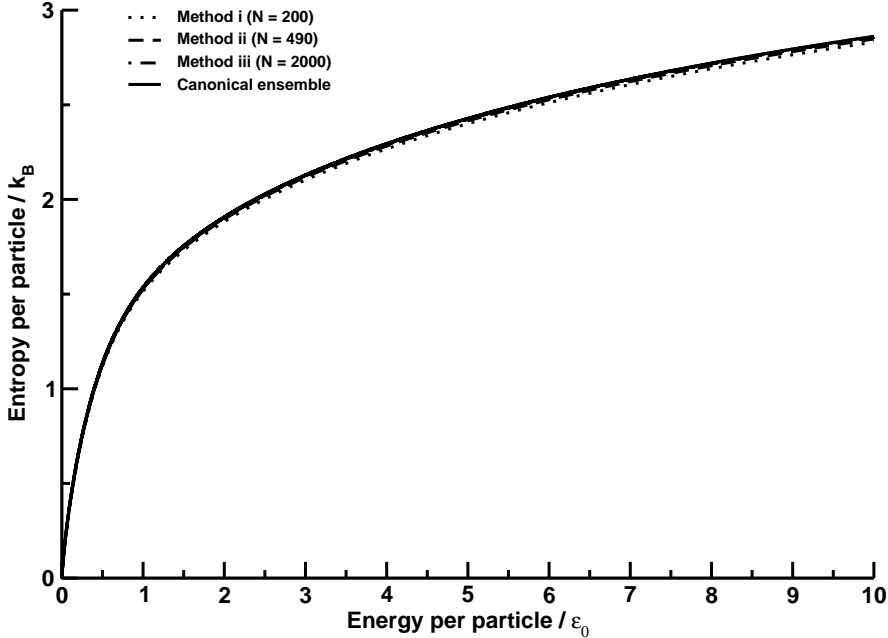}}
\caption{\label{method123}Entropy per particle versus energy per particle calculated for $N=200, 490$ and $2000$ compared with the result calculated in the canonical ensemble ($N\rightarrow\infty$).}
\end{center}
\end{figure}
Figure \ref{method123} shows a plot of entropy per particle versus energy per particle. We notice that the result for $N=200$ is still not converged compared with the result in the infinite system size limit. The deviations for the entropy per particle are of order 1\% at an energy per particle of $10\varepsilon_0$. An improved algorithm must be found.

\subsection{Counting in terms of the smallest real number $\tau$ that is  addressable by the computer\cite{HugeTiny}}
A small improvement in the algorithm can be achieved by counting states in terms of $\tau$, which is the smallest real number that is addressable by the computer rather than in terms of unity which is the normal thing to do. This results in a re-normalized number of states accessible to the system $\tilde{\Omega}(N,E)$ where
\begin{equation}
\tilde{\Omega}(1,E)\,=\,\Omega(1,E)\,\times\,\tau,
\end{equation}
and
\begin{equation}
\tilde{\Omega}(N,E)\,=\,\Omega(N,E)\,\times\,\tau^N.
\end{equation}

The advantage of counting in terms of $\tau$ is that a higher value of $N$ is achievable before $\tilde{\Omega}(N,E)$ exceeds the largest number addressable by the computer. Using this method, the maximum number of particles that we were able to consider is $N\sim 500$. The computational time taken is of the order of minutes. We notice that the result for $N=490$ in Fig.~\ref{method123} is improved over the value for $N=200$, but is still not converged compared with the result in the infinite system size limit. The deviations for the entropy per particle are of order 0.5\% at an energy per particle of $10\varepsilon_0$. In the next subsection, we describe an algorithm that resolves this problem exactly.

\subsection{A recursive solution for the entropy $S(N,E)$}
The solution to this problem comes from the fact that we are ultimately interested in calculating the entropy $S(N,E)$ rather than $\Omega(N,E)$ itself. Rearranging Eq.~(\ref{recursive}) in the following manner
\begin{equation}
\Omega(N,E)\,=\,\Omega(N-1,E)\left(\Omega(1,0)\,+\,\frac{\Omega(N-1,E-1)}{\Omega(N-1,E)}\times\Omega(1,1)\,+\, \cdots\,+\, \frac{\Omega(N-1,0)}{\Omega(N-1,E)}\times\Omega(1,E)\right),
\end{equation}
gives us the following recursive algorithm for the entropy
\begin{equation}
\label{recursive2}
S(N,E)\,=\,S(N-1,E)\,+\,k_B\,\ln\,\sum_{E^{'}=0}^E\,\exp\left(\frac{S(N-1,E-E^{'})-S(N-1,E)}{k_B}\right)\times\Omega(1,E^{'}).
\end{equation}

Working directly with $S(N,E)$ rather $\Omega(N,E)$ is much more manageable from a computational point of view. For the range of $N$ and $E$ that we considered in our work, we never ran into problems exceeding the maximum number addressable by the computer. The computational time taken for $N=2000$ particles and $E=2\times 10^{4}\varepsilon_0$ is of the order of a day. We notice that the result for $N=2000$ in Fig.~\ref{method123} is converged compared with the result in the infinite system size limit. The deviations for the entropy per particle are less than 0.1\% at an energy per particle of $10\varepsilon_0$.

\subsection{Microstate complexity versus averaged macrostate properties}
\begin{figure}[t]
\begin{center}
\scalebox{0.50}{\includegraphics{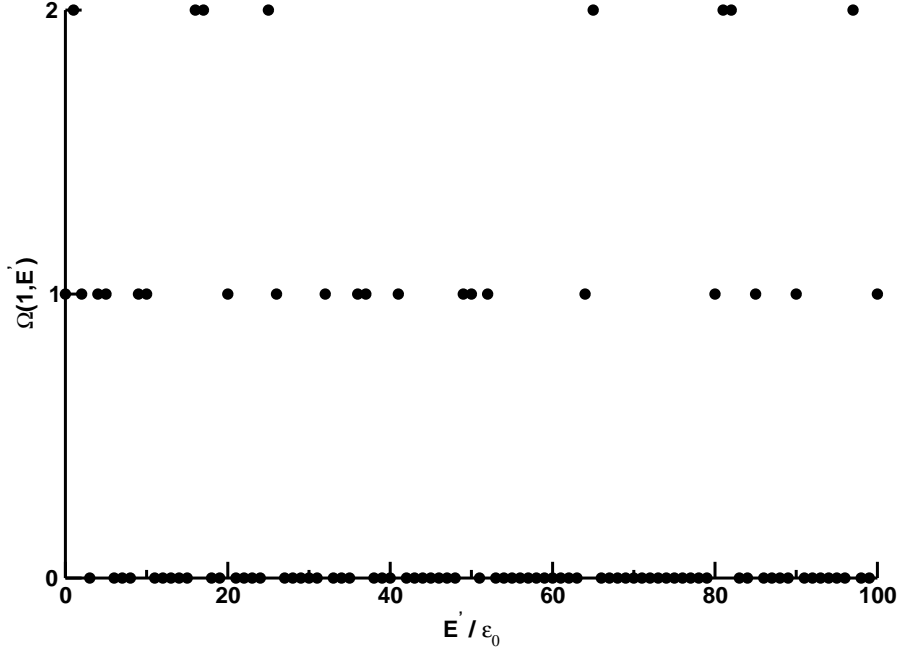}}
\caption{\label{Omega1}The number of microstates $\Omega(1,E^{'})$ for a single particle for $0 \le E{'} \le 100\epsilon_0$. Its discrete structure means that the terms in Eq.~(\ref{recursive}) will be non-zero only when $\Omega(1,E^{'})$ is non-zero.}
\end{center}
\end{figure}
\begin{figure}[t]
\begin{center}
\scalebox{0.50}{\includegraphics{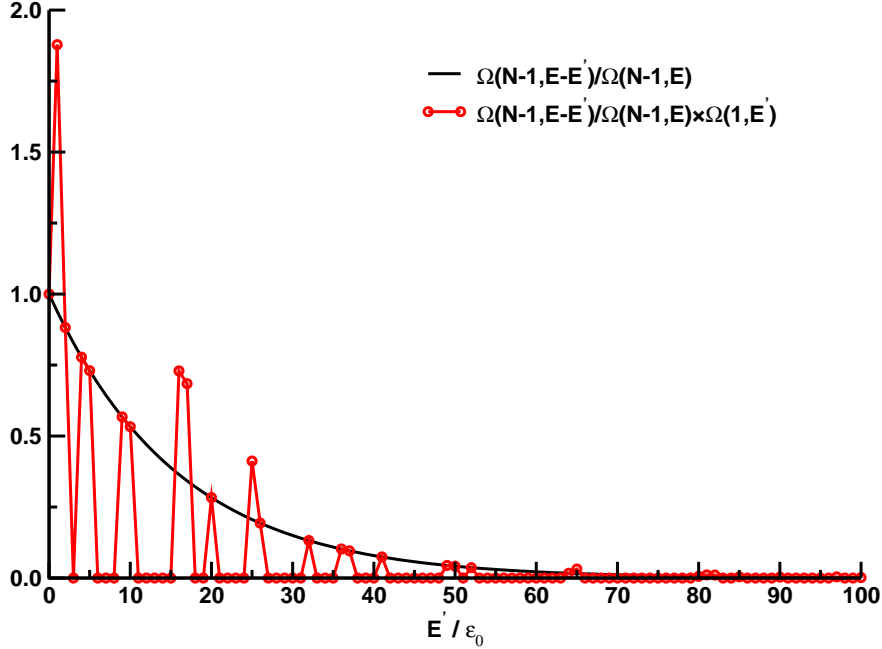}}
\caption{\label{OmegaNminus1} A plot of $\Omega(N-1,E-E^{'})/\Omega(N-1,E)$ versus $E^{'}$ (black curve) and $\left[\Omega(N-1,E-E^{'})/\Omega(N-1,E)\right]\times\Omega(1,E^{'})$ versus $E^{'}$ (red curve with circles) where $0 \le E^{'} \le 2\times10^4\epsilon_0$ and $N=2000$ particles.}
\end{center}
\end{figure}
The complexity of this problem is further illustrated by considering the individual terms in the summation of Eq.~(\ref{recursive}). We have plotted in Fig.~\ref{Omega1}, $\Omega(1,E^{'})$ as a function of $E^{'}$. This is obviously a discrete curve since $\Omega(1,E^{'})$ are the degeneracies of the single particle levels with energy $E^{'}$. We have normalized $\Omega(N-1,E-E^{'})$ by $\Omega(N-1,E)$ for numerical convenience and plotted in Fig.~\ref{OmegaNminus1} for $N=2000$, $\Omega(N-1,E-E^{'})/\Omega(N-1,E)$ and $\left[\Omega(N-1,E-E^{'})/\Omega(N-1,E)\right]\times\Omega(1,E^{'})$ as a function of $E^{'}$ for $E=2\times10^4\varepsilon_0$. Not surprisingly, the curve of $\Omega(N-1,E-E^{'})$ versus $E^{'}$ is smooth and it reflects the convergence in the thermodynamic limit of all intensive macrostate variables such as the entropy per particle, the chemical potential, the temperature and so on.

However, $\Omega(N,E)$ is constructed by summing over terms of the form of the product of $\Omega(N-1,E-E^{'})$ with $\Omega(1,E^{'})$. These contributions are jagged, distinctly discontinuous and rapidly decreasing as a function of $E^{'}$ as seen in Fig.~\ref{OmegaNminus1}. The complexity shown in this figure is non trivial and shows how the microstate structure of the system fundamentally underpins its macrostate properties, even in the thermodynamic limit. No mean-field approach can adequately capture this complexity on the microstate scale.

\subsection{The heat bath}
It is useful to note that the only contributions to $\Omega(N,E)$ come from terms for which $\Omega(1,E^{'})$ are non-zero. So, the subsystem of $(N-1)$ particles can only attain an energy of $E-E^{'}$ if $\Omega(1,E^{'})$ is non-zero. This is true for any value of $N$ and, in particular, this is true as $N\rightarrow\infty$. One can therefore be forgiven for thinking about this as a case of the tail wagging the dog!

However, on more careful inspection of Fig.~\ref{OmegaNminus1}, we note that the main contributions to the summation of Eq.~(\ref{recursive}) come from terms close to $E^{'}=0$ for which the subsystem of $(N-1)$ particles has an energy close to $E$. So, all but the low energy single particle states are suppressed.  In the large $N$ limit, one may view the system of $N$ particles as being composed of one particle in thermal equilibrium with the $(N-1)$ particle subsystem; the latter may therefore be  viewed as a heat bath. This picture enables us to arrive at the Boltzmann distribution that cements the relation with the canonical ensemble. These observations have important implications for the determination of the temperature of the system, which we consider more completely next.

\subsection{Temperature}
\begin{figure}[t]
\begin{center}
\scalebox{0.50}{\includegraphics{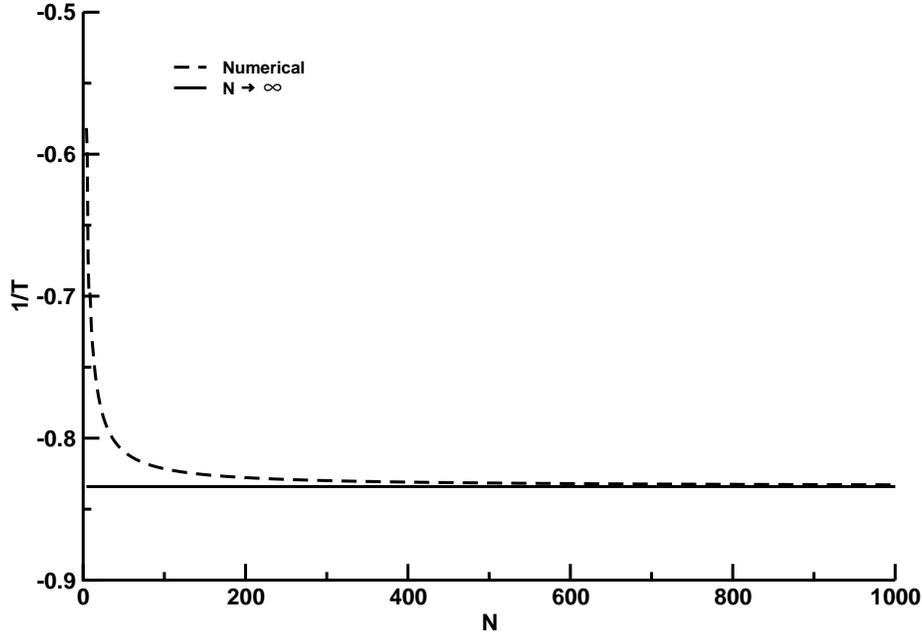}} 
\caption{\label{tempS1}A plot of the inverse temperature as a function of the number of spin $1$ particles, $N$, at fixed energy per particle equal to $0.5\varepsilon_0$. The canonical result ($N \rightarrow \infty$ limit) for the inverse temperature at the same energy per particle is also plotted for comparison. Analytical expressions for thermodynamic quantities within the microcanonical ensemble such as the temperature of the spin $1$ system are usually obtained using the method of Lagrange multipliers.\cite{Pathria,Tuszynski}}
\end{center}
\end{figure}
A simple way to determine the temperature of the system as a function of energy per particle is to apply Eq.~(\ref{Temperature}) to the curve of entropy per particle versus energy per particle in the large $N$ limit. To demonstrate the versatility of our methods we have considered a different model now: We have plotted the temperature of a system of distinguishable spin $1$ particles as a function of $N$ and at energy per particle fixed at $0.5\varepsilon_0$ in Fig.~\ref{tempS1}. In the limit of large $N$ the numerical result converges to the analytical result as it should since these results are derived in the limit of large $N$. The percentage difference between the result in the canonical ensemble ($N \rightarrow \infty$) and the numerical result for $N = 5000$ spin $1$ particles is 0.03\%.

\begin{figure}[t]
\begin{center}
\scalebox{0.50}{\includegraphics{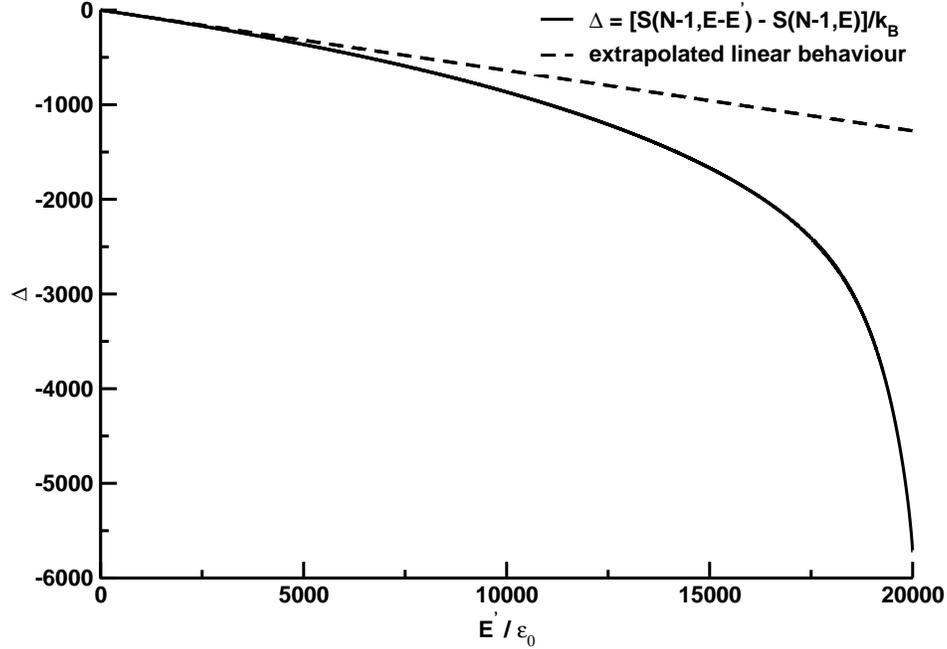}}
\caption{\label{Boltzman}A plot of $\Delta=\,\frac{S(N-1,E-E^{'})\,-\,S(N-1,E)}{k_B}$ as a function of $E^{'}$ for $E=2\times10^4\varepsilon_0$ and for $ 0\leq E^{'} \leq 2\times10^4\varepsilon_0$. The curve is linear for low vales of $E^{'}$. The broken curve in this plot is the linear region extrapolated to $E^{'}=2\times10^4\varepsilon_0$ to show the downward deviation of the curve of $\Delta$ for higher values of $E^{'}$, which corresponds to the super-suppression of the higher energy single particle states. The linear region of the curve of $\Delta$ versus $E^{'}$ has a slope equal to $-0.0638\varepsilon_0^{-1}$, which we identify as $-\beta$ according to Eq.~(\ref{Beta}).}
\end{center}
\end{figure}

A more interesting way to extract the temperature that cements the equivalence of the microcanonical ensemble with the canonical ensemble is now presented. Following Eq.~(\ref{recursive2}), and for $N=2000$, which we have already demonstrated is sufficiently close to the thermodynamic limit, we have plotted in Fig.~\ref{Boltzman},\begin{equation}\Delta\,=\,\frac{S(N-1,E-E^{'})\,-\,S(N-1,E)}{k_B}\end{equation} as a function of $E^{'}$ for $E=2\times10^4\varepsilon_0$.

The results are negative, as they should be, and linear for low values of $E^{'}$. The curve deviates downwards, i.e. becomes more negative, for higher values of $E^{'}$ which corresponds to the super-suppression of the higher energy single particle states. This super linear behaviour for higher values of $E^{'}$ we found to be very intriguing and we are not aware that this has been noted before. For the low energy single particle states, we may adequately model this curve by
\begin{equation}
\label{Beta}
\Delta\,=\,-\beta\,E^{'}
\end{equation}
so that Eq.~(\ref{recursive2}) may now be re-written as
\begin{equation}
\label{recursive3}
S(N,E)\,=\,S(N-1,E)\,+\,k_B\,\ln\,\sum_{E^{'}=0}^{\infty}\,\exp (-\beta\,E^{'})\times\Omega(1,E^{'}).
\end{equation}
This works fine for the higher energy states as well which are suppressed even further than what the linear relation in Eq.~(\ref{Beta}) implies. The upper limit in the summation in Eq.~(\ref{recursive3}) has therefore, accordingly, been replaced with $\infty$ with no change to the final result. $\beta$, of course, is a function of $E$ and emerges here simply from the linear fit to the low energy data in Fig.~\ref{Boltzman}. This establishes the equivalence with the canonical ensemble for which $\beta$ is identified with the inverse temperature \begin{equation}\beta\,=\,\frac{1}{k_BT}\end{equation} and $z(T)$ with the single particle partition function \begin{equation}z\,=\,\sum_{E^{'}=0}^{\infty}\,\exp (-\beta\,E^{'})\times\Omega(1,E^{'}).\end{equation}

\subsection{Chemical potential}
Since Eq.~(\ref{recursive}) corresponds to us systematically increasing the system size by one particle at a time, a very natural thermodynamic quantity to consider is the chemical potential which is defined as
\begin{equation}
\mu\,=\,\left(\frac{\partial E}{\partial N}\right)_{SV}.
\end{equation}
Since it is very difficult to numerically keep the entropy fixed (the preceding sections will attest to this!), it is more useful for us to consider the quantity
\begin{equation}
\label{ChemicalPotential1}
-\frac{\mu}{T}\,=\,\left(\frac{\partial S}{\partial N}\right)_{EV},
\end{equation}
instead. Because $\delta N=1$, we may write
\begin{equation}
\label{ChemicalPotential2}
-\frac{\mu}{T}\,=\,S(N,E)-S(N-1,E),
\end{equation}
which can be easily evaluated using Eq.~(\ref{recursive2}).

Plotting $-\frac{\mu}{T}$ as a function of $N$ will give an indication of the convergence of $-\frac{\mu}{T}$ in the thermodynamic limit. However, there is a subtle point that is worth noting. The differentiation process in Eq.~(\ref{ChemicalPotential1}) requires that the total energy $E$ must be kept constant. However, if we keep $E$ constant in Eq.~(\ref{ChemicalPotential2}), then $E/N\rightarrow 0$ as $N\rightarrow \infty$. In this case, the entropy per particle $\rightarrow 0$ as $N\rightarrow \infty$, and Eq.~(\ref{ChemicalPotential2}) simply converges to zero. This is correct but not very useful! It is more instructive to keep the energy per particle fixed in Eq.~(\ref{ChemicalPotential2}), and in so doing the correct convergence properties of $-\frac{\mu}{T}$ will be achieved. We leave this as an exercise for the reader to investigate.

A second subtle point related to the chemical potential as defined in Eq.~(\ref{ChemicalPotential2}) is that the left hand side of this equation is manifestly intensive, whereas the right hand side is manifestly extensive. The reader is encouraged to resolve this dilemma. The results can be investigated numerically according to the discussion in the previous paragraph.

\begin{figure}[t]
\begin{center}
\scalebox{0.50}{\includegraphics{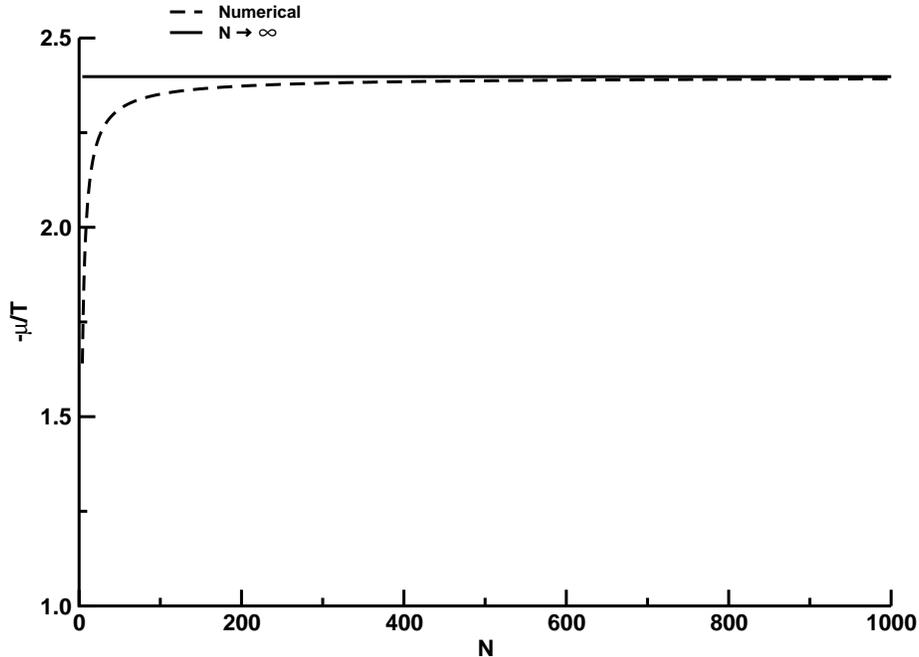}} 
\caption{\label{ChempotHO}A plot ($-\mu/T$) as a function of the number of harmonic oscillators, $N$, at fixed energy per particle equal to $10\varepsilon_0$. The canonical result ($N \rightarrow \infty$ limit) for the chemical potential at the same energy per particle is also plotted for comparison. The analytical expression for the inverse temperature as a function of the energy per particle, $e$, is $-\mu/T = k_B \ln(e+1)$.}
\end{center}
\end{figure}

Another useful way to approach an investigation of the chemical potential is to consider the following scaling argument. Let $s(e)$ be the entropy per particle in the thermodynamic limit, where $e$ is the energy per particle. Then
\begin{equation}
S(N,E)\,=\,N\,s(\frac{E}{N}).
\end{equation}
Applying Eq.~(\ref{ChemicalPotential1}) to the above expression gives
\begin{equation}
\label{mu2}
-\frac{\mu}{T}\,=\,s(e)\,-\,e\frac{d\,s(e)}{de}.
\end{equation}
Once again, to demonstrate the versatility of our method, we have plotted this expression in Fig.~\ref{ChempotHO} for a system of quantum harmonic oscillators as a function of $N$ to test the convergence of $-\frac{\mu}{T}$ in the thermodynamic limit. This is done at constant energy per particle $e$ equal to $10\varepsilon_0$ rather than constant total energy $E$. The numerical result for the chemical potential calculated using Eq.~(\ref{mu2}) converges to the analytical result in the limit of large $N$. The percentage difference between the analytical result (in the limit of $N \rightarrow \infty$ ) and the numerical result for $N=2000$ harmonic oscillators is 0.1\%. 

\section{Final Remarks}
We have developed a hierarchy of methods that enable the computation of the entropy for a system of independent and distinguishable particles. Our methods are very general and applicable to a very wide class of problems. Our algorithm expressed in Eq.~(\ref{recursive2}) is robust and may be studied in the large $N$ limit with a reasonable computational effort. To within limits set by the computer, $N$ can be made arbitrarily large. This method enables one to monitor the approach to the thermodynamic limit as a function of $N$ for various thermodynamic quantities such as the entropy per particle, the temperature, the chemical potential, and so on. 

We have given some exercises for the reader to pursue this subject more completely. It is helpful to first apply our expressions to an analytically solvable model such as a system of independent quantum harmonic oscillators or a system of independent spin $1$ particles. Some of the arguments, for example those referred to the chemical potential in Section II J, may first be tested against known analytical results.

Considering indistinguishable spinless particles (bosons) or indistinguishable spin $\frac{1}{2}$ particles (fermions) would be a natural extension to our methods.

\appendix
\section{The canonical ensemble}
In the text, results for various properties in the microcanonical ensemble for finite $N$ are compared with equivalent results in the canonical ensemble which are, by definition, in the thermodynamic limit. The relevant expressions for the canonical ensemble are given below.

The single-particle partition function $z(X)$ for the model described in Eq.~(\ref{energy}) is given by
\begin{equation}
z(X)\,=\,\sum_{p=0}^{\infty}\exp(-Xp^2)\,\times\,\sum_{q=0}^{\infty}\exp(-Xq^4),
\end{equation}
where $p$ and $q$ are non-negative integers, and where \begin{equation}X=\frac{\varepsilon_0}{k_BT}\end{equation} is dimensionless and is a measure of the inverse temperature. $z(X)$ as a function of $X$ can be computed numerically in a very straight-forward manner.

From this, the Helmholtz free energy per particle $f(X)$ in units of $\varepsilon_0$ is given by
\begin{equation}
f(X)\,=\,-\frac{1}{X}\ln z(X),
\end{equation}
and the average total energy per particle $e(X)$ in units of $\varepsilon_0$ is given by
\begin{equation}
e(X)\,=\,-\frac{d}{dX}\ln z(X).
\end{equation}

We calculate the average entropy per particle in units of $k_B$ from
\begin{equation}
s(X)\,=\,X\,(e(X)\,-\,f(X)).
\end{equation}
This enables us to plot, for example, the entropy per particle versus the energy per particle (at the same value of $X$).

Another useful result for our analysis is the expression for the chemical potential in units of $\varepsilon_0$ which is given by
\begin{equation}
\mu\,=\,f(X).
\end{equation}


\begin{thebibliography}{}
\bibitem{HugeTiny}  For our 32-bit computer, the largest real number addressable by the computer is 1.797693134862316E+308, and the smallest real number is 2.225073858507201E-308.
\end{thebibliography}
\end{document}